# Sb$_2$S$_3$ and GaAs Absorber Layer-based Quantum Dot Solar Cells with Cadmium Telluride-based HTL: A Comparative Study

*Sayak Banerjee[1,2], Anupam Chetia[1], Satyajit Sahu[1]*

[1]*Department of Physics, Indian Institute of Technology, Jodhpur, Rajasthan, India-342037*

[2]*Department of Metallurgical and Materials Engineering, Indian Institute of Technology, Jodhpur, Rajasthan, India-342037*

**Abstract:** Quantum dot solar cells (QDSC) are widely acknowledged to be one of the best solar energy harvesting devices in the present world. Absorber layer is a core component of a QDSC with a strong influence on its operational efficiency. Hence, we choose to undertake a comparative study of two QDSC having different QD absorber layers: Sb$_2$S$_3$ and GaAs with the motive to identify the better absorber layer material. The numerical analysis has been carried out using SCAPS-1D (Solar Cell Capacitance Simulator-1D). The structure of the QDSCs under study are: FTO/TiO$_2$/CdS/Sb$_2$S$_3$/CuI/C and FTO/TiO$_2$/CdS/GaAs/CuI/C. Critical parameters, including temperature, back contact work function, series and shunt resistances, were meticulously adjusted in the simulations, demonstrating that the maximum efficiency attained by Sb$_2$S$_3$ and GaAs absorber layer based QDSC is 15.94% and 26.95% respectively indicating GaAs-QD to be a better absorber layer material for a QDSC.

# 1. Introduction

The conventional sources of energy will not be sufficient enough to fulfil the increasing demands of power in the near future and are responsible for the rapid increase in greenhouse gases leading to climate change [1]. These sources being non-renewable are not likely to be replenished once exhausted, hence the need to develop an alternative, economical and sustainable source of energy, having negligible environmental impact was felt and are presently under active research [2]. Out of the various replenishable sources of energy available on Earth, solar energy is the dominant one. Consequently, there arose a demand for the extraction of solar energy, leading to the emergence of solar photovoltaic technology. Solar cells operate based on the principles of photovoltaic effect to convert solar energy into consumable electricity. Solar cell technology can be broadly classified into three categories: The 1st generation solar cell or the conventional Silicon solar cells accounting for more than 80% of the commercial solar panels sold world-wide [3]. Second-generation solar cells, or the thin-film solar cells, consist of semiconductor materials arranged in multiple layers with a thickness typically measuring a few microns. This particular type of solar cell has a remarkably low manufacturing cost and have acquired an impressive efficiency of nearly 20% [4]. The third-generation solar cells are composed of a diverse variety of innovative materials [5] such as quantum dots, nanotubes, nanowires, organic dyes and so on. QDSC (quantum dot solar cell), perovskite solar cell (PSC), dye sensitized solar cell (DSSC) and nanocrystalline solar cell are 3rd generation solar cells having excellent properties that are capable of achieving high efficiency and are areas of active research today.

Quantum dot solar cell have outstanding optical and electronic properties with tunable band gap [6] that fulfil our requirement of converting a wider band of solar spectrum into consumable electricity so to improve the efficacy of the solar cell. They have the ability of achieving a maximum efficiency of 66% (theoretically) considering infinite stack of cascaded multiple p-n junctions [7] and multiple exciton generation, which is more than twice as that of the efficiency of the present day commercially available solar cells.

In this paper, we will carry out the optimization and comparative study of two different configuration of solar cell having different quantum dot absorber layer. In one of the configurations, we have $Sb_2S_3$ quantum dot while in the other we have GaAs quantum dot as the absorber layer. To ensure an accurate comparison, the doping concentration and thickness of both absorber layers will remain constant. We have set AM 1.5G with power 1000 watt/m$^2$ as the solar illumination and temperature 300K throughout the simulation of the QDSC. Finally, the optimised solar cells are subjected to unavoidable series and shunt resistances, to find out the maximum efficiency achievable by these solar cells. It has been found after optimization that the GaAs-QD based solar cells excels the other by an efficiency of more than 10% achieving a remarkable efficiency of 26.95% in presence of presence of parasitic resistance.

## 2. Numerical simulation technique

The solar simulating device SCAPS-1D, developed by ELIS, University of Gent is used by us for the simulation of the QDSC. The device allows designing of up to seven semiconductor layers to perform the simulation. We get the liberty to re-define some extrinsic properties of the semiconductor materials used. We can also introduce defects in the layers and also in between the layers at the interfaces. Discretion is available to vary the properties of the defects introduced. There are provisions to vary working temperature, resistances, spectrum of solar illumination and study the effects due to their variation.

All the solar cell characterization parameters such as Voc, Jsc, FF, PCE, quantum efficiency, energy bands are evaluated directly using this simulating device [8].

The simulation device computes these parameters by solving coupled differential equations namely the Poisson equation (1), the hole continuity equation (2) and the electron continuity equation (3) at provided boundary conditions [9][10].

$$\frac{\partial}{\partial x}\left(-\epsilon(x)\frac{\partial V}{\partial x}\right) = q[p(x) - n(x) + N_D^+(x) - N_A^-(x) + p_t(x) - n_t(x) \quad (1)$$

$$\frac{\partial p}{\partial t} = -\frac{1}{q}\frac{\partial J_p}{\partial x} + G_p - R_p \quad (2)$$

$$\frac{\partial n}{\partial t} = \frac{1}{q}\frac{\partial J_n}{\partial x} + G_n - R_n \quad (3)$$

Where $\epsilon, V$ denotes dielectric permittivity of the material and electric potential respectively, whereas $q, p(x), N_A^-(x), p_t(x), n(x), N_D^+(x), n_t(x), J_{n,p}, G_{n,p}, R_{n,p}$ denotes the electronic charge, free hole concentration, ionized acceptor concentration, trap density of holes, concentration of free electron, ionized donor concentration, and trap density of electrons, electron-hole current density, electron-hole generation rate and electron-hole recombination rate respectively [9].

## 3. Quantum dot solar cell structure

The QDSC comprises three principal layers namely electron transport layer (ETL) – responsible for electron separation and transportation to the respective electrode, tunable band gap quantum dot absorber layer for solar radiation absorption and subsequent generation of exciton and the hole transport layer (HTL) – responsible for hole separation and transportation to the respective electrode [11][12][13]. Apart from this an intermediate layer at the junction of ETL-absorber layer called the buffer layer may also be present. The absorber layer when remain coupled with a buffer layer, absorbs huge amount of solar radiation thereby plays an important role in maximizing the effectiveness of a solar cell [8], [11] These layers are enclosed by two contacts on the front and the rear end of the solar cell. Generally, transparent conducting oxides such as ITO (Indium doped Tin Oxide), FTO (Fluorine doped Tin Oxide) are employed as the front contact. Materials such as Copper (Cu), Gold (Au), Silver (Ag), Carbon(C), Platinum (Pt), Nickel (Ni) are generally used as the back contact [13].

We have choosen $TiO_2$ as the ETL because as per knowledge from previous researches which showed that out of various other materials available $TiO_2$ is capable to yielding better results due to its appropriate bandgap (3.26 eV) [13], high mobility of electron, eco-friendliness and low cost of synthesis [14]. Buffered between the ETL and the QD-Absorber layer lies the CdS layer, which is coupled with the absorber layer for better absorption of sunlight. Generally, an n-type material is suitable for a buffer layer and CdS being an n-type material with high band gap (2.4 eV) is a perfect match as a buffer layer. The layer following the buffer layer is the quantum dot-based absorber layer. The two QDSC differ from each other due to this layer. In one of the QDSC we have used $Sb_2S_3$ quantum dot while in the other we have used GaAs QD. It has been reported in various research works that quantum dots having band gap within the range of 0.76-1.6 eV exhibit good efficacy [15] and both the quantum dots we have used have their band gaps in this range.

$Sb_2S_3$ is a highly promising contender for the absorber layer because of its elevated absorption coefficient, matching bandgap (1.6-1.8 eV), single stable phase and suitable synthesis conditions - (can be synthesized at temperatures below $350°C$). Apart from this, $Sb_2S_3$ has low melting point (550 °$C$) and has achieved maximum theoretical efficiency of about 40% as per previous research reports [16].

GaAs having band gap 1.43 eV is an excellent solar cell absorber material. Previously recorded work with GaAs includes thin film solar cell where GaAs is the absorber material and has PCE as high as 19.9% and 22.08% [17][18].

A variety of materials can be used as the HTL but it must be noted that it must align well with the quantum dot absorber layer of the QDSC. Various research reports claim CuI to be a good candidate for the HTL. Its low electron affinity [19], high hole mobility [20] are reasons for the same. The solar cell structure completes with the back contact or the anode. Materials those can be considered to be the back contact are Silver, Carbon, Gold, Nickel, Platinum etc. Carbon being very cost effective and provide good results, hence we will use Carbon as the back contact in our solar cell configuration.

Two quantum dot solar cell are thus ready for characteristic study, one being FTO/TiO$_2$/CdS/Sb$_2$S$_3$/CuI/Carbon while the other being FTO/TiO$_2$/CdS/GaAs/CuI/Carbon

Figure 1 (a) and (b) shows the quantum dot solar cell structures to be used to carry out the comparative study between the two and their corresponding band energy structures represented by figure (c) and (d) respectively.

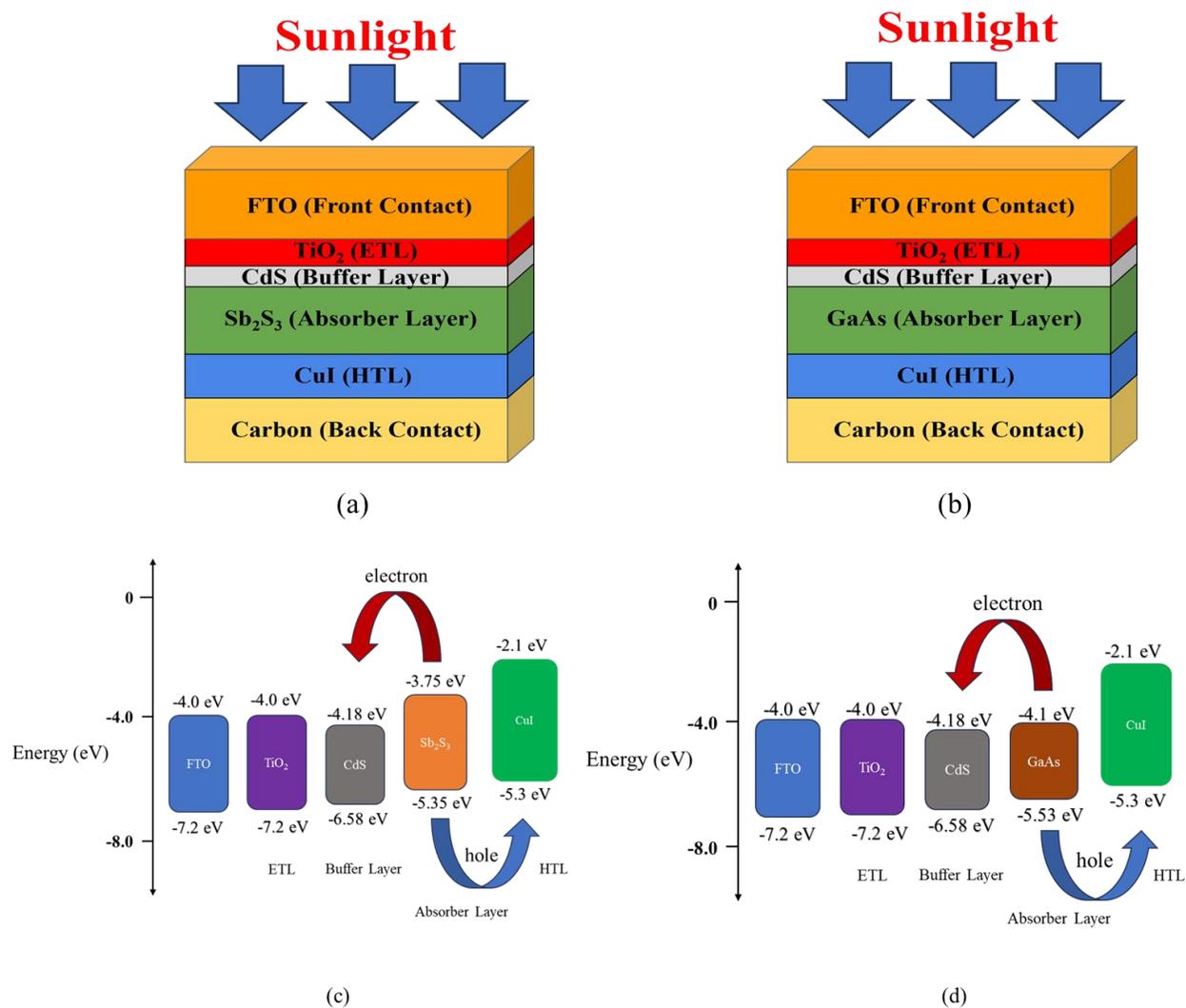

**Figure 1** Representing the configurations of the quantum dot solar cell having (a) $Sb_2S_3$ and (b) GaAs quantum dot as the absorber layer and their corresponding band energy structures represented in (c) and (d) respectively.

Table-1 displays the characteristic properties of the materials used in the simulation.

**Table-1:** Characteristic properties of the materials used in the device.

| Material Properties | FTO | TiO$_2$ | CdS | Sb$_2$S$_3$ | GaAs | CuI |
|---|---|---|---|---|---|---|
| Thickness (μm) | 0.45 | 0.075 | 0.02 | 0.40 | 0.40 | 0.15 |
| Eg (eV) | 3.20 | 3.26 | 2.40 | 1.60 | 1.43 | 3.2 [21] |
| X (eV) | 4.40 | 4.20 | 4.2 | 3.70 | 4.07 | 2.1 [22] |
| $\epsilon$ | 9.0 | 10.0 | 9.0 | 7.08 | 12.9 | 6.5 [22] |
| N$_C$ (cm$^{-3}$) | $2.2 \times 10^{18}$ | $2.0 \times 10^{17}$ | $1.0 \times 10^{18}$ | $2.63 \times 10^{19}$ | $4.7 \times 10^{17}$ | $1.92 \times 10^{18}$ |
| N$_V$ (cm$^{-3}$) | $1.8 \times 10^{19}$ | $6.0 \times 10^{19}$ | $1.0 \times 10^{19}$ | $6.25 \times 10^{19}$ | $9.0 \times 10^{18}$ | $2.24 \times 10^{18}$ |
| N$_A$ (cm$^{-3}$) | 0 | 0 | 0 | $1.0 \times 10^{15}$ | $1.0 \times 10^{15}$ | $1.0 \times 10^{19}$ |
| N$_D$ (cm$^{-3}$) | $2.0 \times 10^{19}$ | $1.0 \times 10^{18}$ | $1.0 \times 10^{17}$ | 0 | 0 | 0 |
| μ$_e$ (cm$^2$/Vs) | 20 | 100 | 50 | 50 | 8500 | 48.78 |
| μ$_h$ (cm$^2$/Vs) | 10 | 25 | 20 | 10 | 400 | 43.90 [20] |
| Density of defects (cm$^{-3}$) | $1.0 \times 10^{14}$ | $1.0 \times 10^{14}$ | $1.0 \times 10^{14}$ | $1.0 \times 10^{14}$ | $1.0 \times 10^{14}$ | $1.0 \times 10^{14}$ |
| Energetic distribution | Single | Single | Single | Single | Single | Single |
| Defect nature | Neutral | Neutral | Neutral | Neutral | Neutral | Single Donor |
| Capture cross-section of holes (cm$^2$) | $10^{-16}$ | $10^{-16}$ | $10^{-12}$ | $10^{-17}$ | $10^{-17}$ | $10^{-15}$ |
| Capture cross-section of electrons (cm$^2$) | $10^{-16}$ | $10^{-16}$ | $10^{-17}$ | $10^{-17}$ | $10^{-17}$ | $10^{-12}$ |
| Defect energy level with respect to reference (eV) | 0.600 above EV | 0.600 above EV | 1.200 above EV | 0.600 above EV | 0.600 above EV | 0.750 above EV |
| References | [9] | [13] | [23] | [23] | [13] | [24] |

## 4. Results and analysis

### 4.1 Absorber layer optimization in both QDSC

In the course of optimization of the absorber layer in both the QDSC, we have focussed on to three aspects, the thickness, the concentration of acceptor dopants in the layer and the density of defects present in the layer.

### 4.1.1 Comparative analysis of impact of thickness of absorber layer on both QDSC

A solar cell efficiency is highly determined by the thickness of its absorber layer. The thickness must be so chosen to generate a large current but it should not be too large to create negative impact and reduce the efficiency [25] The absorber layer plays a crucial role in the absorption of photons and generation exciton for the conduction of electricity in a solar cell. The density of photon absorption by the absorber layer depends largely on the thickness of a layer. A thicker layer is capable of absorbing large number of solar photons incident on it. It must be noted that a very thick absorber layer although absorbing huge number of photons, it also increases the chances of e-h recombination thereby lowering the power output of the solar cell. Therefore, it is necessary to have an optimal thickness of the absorber layer in order to ensure that the absorber layer is neither too thick to induce electron hole recombination nor too thin that it doesn't absorb enough photons required for generation of power.

Based on previous experimental works reported, it has been seen that researchers have limited the thickness of absorber layer to less than 600 nm even when sometimes they got better results at higher thickness. This is because of fabrication and characterization challenges faced in laboratory. Hence, we too during our simulation have restricted our study within the range of 350-550 nm for both the configuration of the QDSC.

Figure-2 represents the comparative analysis of the impact of thickness of the QD absorber layer on (a) current density and (b) PCE of the two QDSC.

In both the configuration, the current density and efficiency is seen to rise attaining the maximum value at a thickness of 550 nm. Attaining a high current density of 28 mA/cm$^2$ and an efficiency ranging from 22.81-25.42%, the GaAs QDSC is found to be dominating in comparison to the Sb$_2$S$_3$ QDSC having current density of 22.2 mA/cm$^2$ and an efficiency ranging between 13.03-14.70 % when the thickness of the absorber layer is raised within the range of 350-550 nm in equal steps. Hence, the optimised thickness of the absorber layer is 550 nm for both the QDSC. The 10% difference in efficiency of the two QDSC at similar conditions may be attributed to the different QD material used in the absorber layer. Sb$_2$S$_3$ has higher band gap as compared to GaAs, this causes reduced absorption of light by Sb$_2$S$_3$ QD in its energy range. Hence, GaAs-QD with an optimal band gap of 1.43 eV is far ahead of Sb$_2$S$_3$ QD in terms of efficiency and current generation.

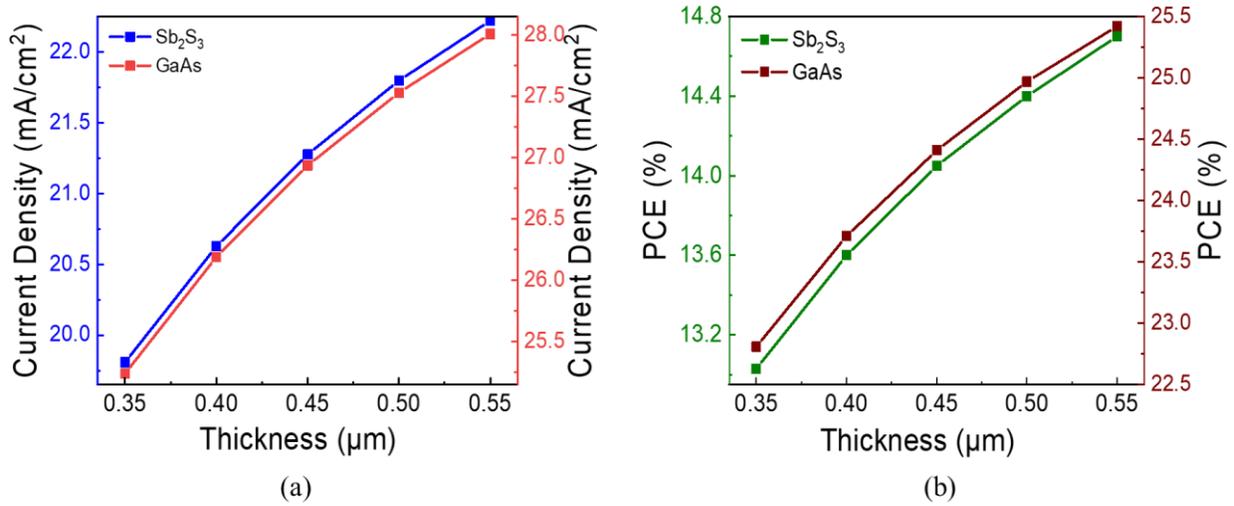

**Figure 2** Comparative study of effect of thickness of the QD absorber layer of the two QDSC on (a) Current Density (b) Power conversion efficiency.

### 4.1.2 Comparative analysis of impact of acceptor doping concentration of absorber layer on the two QDSC

Along with thickness, doping concentration (acceptor/donor) is no less an important parameter that controls the efficacy of solar cells. The concentration of acceptor dopants may affect the electrical properties of a material such as its carrier mobility, conductivity, etc and thereby likely to create an overall effect on the efficiency of solar cells.

Here the density of defects in the QD-absorber layer is made to vary within the range of $10^{14} - 10^{19}\ cm^{-3}$ [26] and it has been noticed that with the rise in concentration of acceptor doping of the QD-absorber layer, the photo-current density increases and stabilizes nearly at $10^{19}$ cm$^{-3}$ for both the QDSC devices. With regard to the efficiency of power conversion, initially the efficiency reduces to reach the lowest at $10^{17}$ cm$^{-3}$ beyond which the efficiency increases when the doping concentration is further increased. However, studying the combined effect of current density and the power conversion efficacy, it has been seen that $10^{19}$ cm$^{-3}$ is the optimized level of concentration of acceptor dopants in the QD-absorber layer in both QDSCs.

Figure-3 represents the effect on (a) current density (b) power conversion efficiency due to variation in density of defects in the absorber layer.

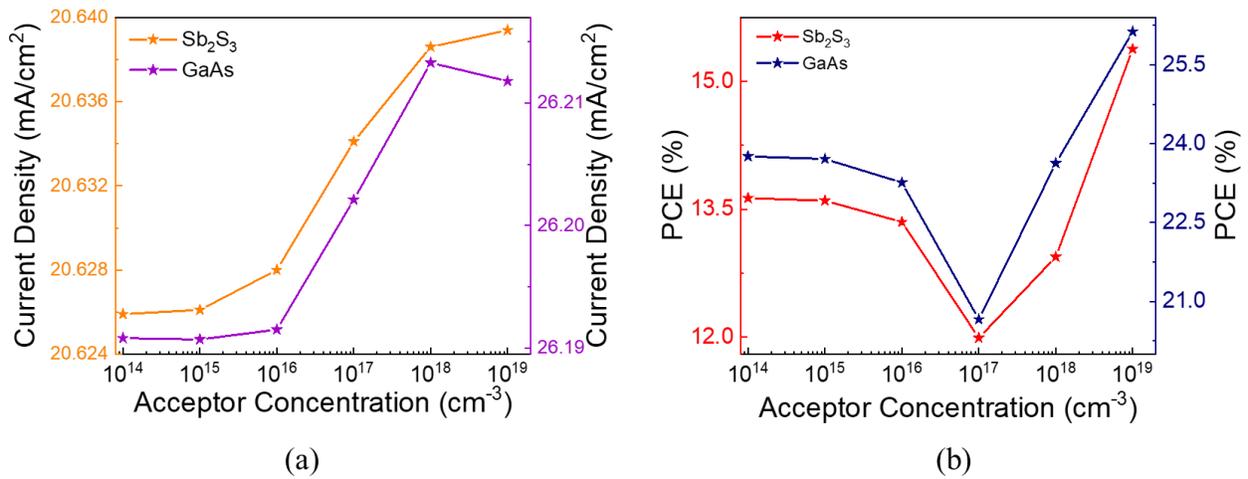

**Figure 3** Comparative study of impact of thickness of the QD absorber layer of the two QDSC on (a) Current Density (b) Power conversion efficiency.

### 4.1.3 Comparative study of effect of defect density of absorber layer in both QDSC

While we deal with various materials, it is highly possible that the material may get subjected to some undesirable defects. These defects may be in various forms- point like defects such as Frenkel or Schottky or may be vacancies or interstitials. They may also be in form of grain boundaries or dislocations [27]. The more the density of defects, the higher are the chances of recombination of the photo-generated carriers. Hence tends to lower the solar cell efficacy [28].

Here the density of defects of the QD-absorber layer is made to vary within the range of $10^{14} - 10^{19}\ cm^{-3}$. Figure-3 represents the effect on (a) current density (b) power conversion efficiency due to variation of defect density in the QD-absorber layer.

It has been noticed that for both the QDSC, the current density and hence the power conversion efficiency decreases with rise in defect density following the normal trend as discussed above. In case of photovoltaic current density, the decrease in both the QDSC is almost same. Neither of them is found to be dominating against the other. The decrement is from 20.63-20.44 mA/cm² in case of $Sb_2S_3$ QD based solar cell, while 26.19-26.18 mA/cm² for the GaAs QD based solar cell. In case of effect on overall efficacy of solar cell, it has been observed that GaAs based solar cell suffers a huge setback in contrast to the $Sb_2S_3$ QDSC. In the former one mentioned, efficiency decreases from 23.71-17.7 %, whereas in the latter, the efficiency drop is not significant decreasing from 13.6-12.6 %. The significant drop in efficiency in case of GaAs QD based solar cell may be attributed to narrow band gap of GaAs in comparison to Sb$_2$S$_3$ as narrow band gap materials are more sensitive towards defects. Hence, we have set $10^{14} cm^{-3}$ as the optimised density of defects for the absorber layer in both the QDSC.

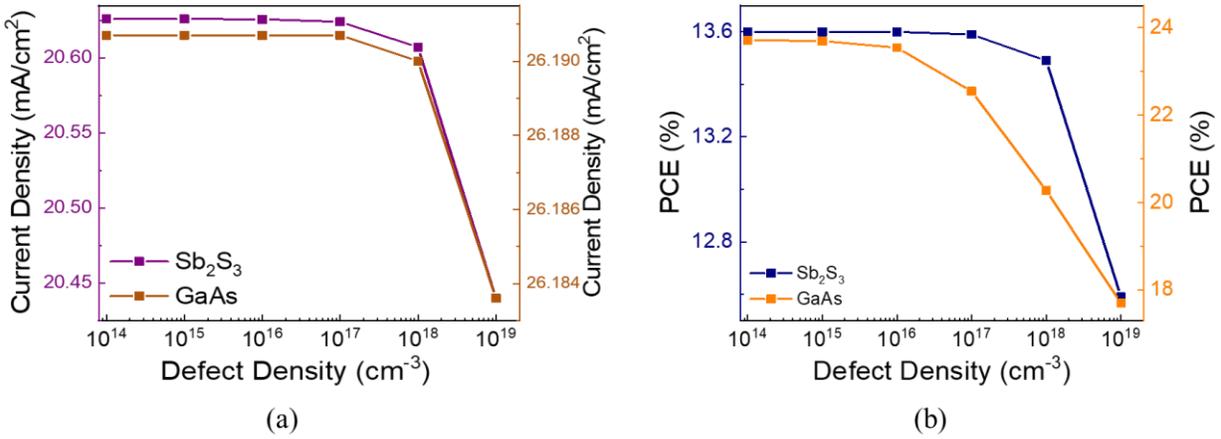

**Figure 4** Comparative study of impact of defect density of the QD absorber layer of the two QDSC on (a) Current Density (b) Power conversion efficiency.

## 4.2 Optimization of ETL

In this sub-section, we will deal with the ETL optimization in both the QDSC. The optimization is carried out in two aspects, the optimization of thickness and doping concentration.

### 4.2.1 Comparative study of effect of thickness of ETL of both the QDSC

Initially both the QDSC had same ETL thickness. The ideal thickness of electron transport layer should be very thin. However, it is not always possible to reduce it significantly due to practical constraints [29]. To find out whether the thickness of ETL that we have set initially is capable of providing high efficiency, we will vary the ETL thickness in both the solar cell ranging from 50-100 nm in intervals of 10 nm and check its effect on current density and the solar cell's electrical power conversion efficiency.

Figure-5 illustrates the impact of TiO2 ETL thickness of Sb$_2$S$_3$ and GaAs QD based solar cell on (a) photovoltaic current density (b) power conversion efficiency.

It has been observed that the current density gradually decreased as the ETL thickness raised, for both the QDSC. Decrement of current density is also indicative of the fact that PCE would decrease, provided $V_{oc}$ decrease too with increase in ETL thickness. The same ETL material being used in both the QDSC, they show a similar trend. However, due to the absorber layer being different, both have a difference in their efficiency ranges. Thus, we have reduced the ETL thickness to 75 nm to 50 nm in order to enhance the efficiency of the solar cell. The thinner ETL results in better efficiency as in case of thinner ETL, photocurrent generation is higher due to the shorter distance travelled from absorber layer to the front electrode [29].

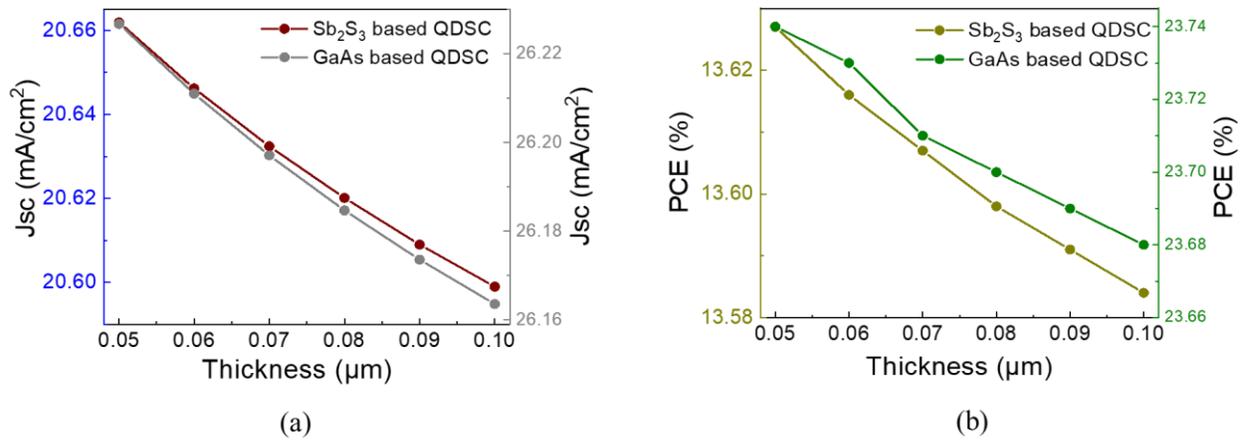

**Figure 5** Comparative study of impact of $TiO_2$ ETL thickness of the two QDSC on (a) Photovoltaic current density (b) Efficiency of the solar cell.

### 4.2.2 Comparative study of effect of concentration of donor dopants in ETL of both the QDSC

This study compares how the donor concentration of ETL impacts the current density and power conversion efficacy of both QDSCs. Initially both the solar cell had ETL donor doping concentration of $10^{18}\ cm^{-3}$. To maximize efficiency, it is necessary to have the concentration of the donor dopants to the optimal level. For the same, the concentration of donor dopants in the ETL in both the QDSC is varied from $10^{15} - 10^{19}\ cm^{-3}$.

Figure-5 illustrates the fluctuation of (a) photovoltaic current density (b) power conversion efficiency with increase in donor concentration of ETL.

It has been noted that with the rise in concentration of the donor dopants in the ETL, the current density starts decreasing slowly in the beginning and then the fall is sharp in the case of both QDSC. The $J_{sc}$ values decreases from 20.73-20.52 mA/cm$^2$ in Sb$_2$S$_3$ QDSC and from 26.30-26.09 mA/cm$^2$ in GaAs QDSC. This decrease in current density with rise in donor doping concentration may be attributed to the decrease in carrier mobility as a result of charge-carrier recombination [30]. On the other hand, PCE is seen to rise with increase in donor concentration of the ETL. PCE increases from 13.22-13.63 mA/cm$^2$ for the Sb$_2$S$_3$ QDSC and from 23.20-23.74 mA/cm$^2$ for the GaAs QDSC.

We therefore, fix $10^{19}\ cm^{-3}$ as the donor concentration in both the QDSC as it is giving the maximum efficiency in case of both the QDSC.

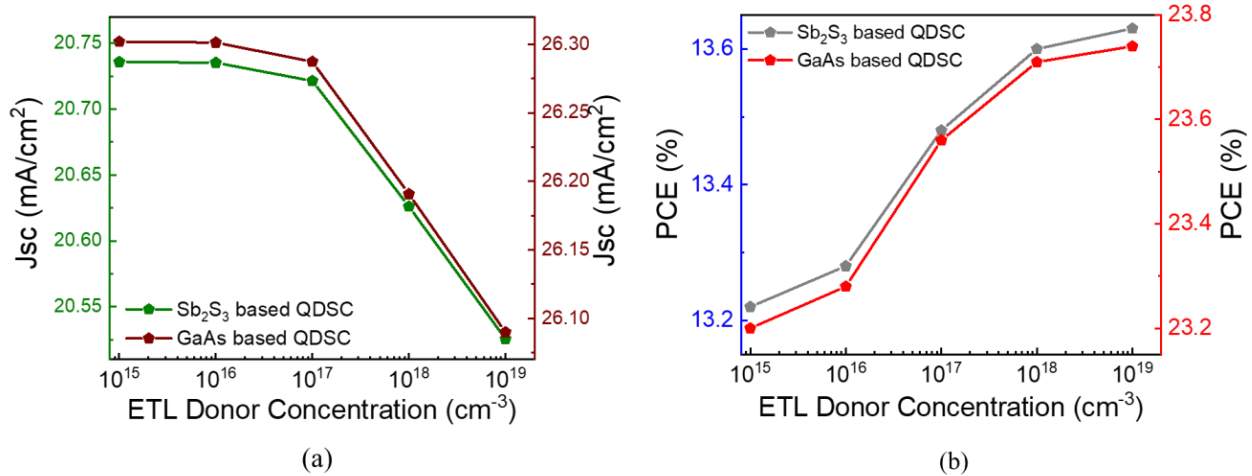

**Figure 6** Comparative study of ETL donor concentration impact on (a) Current Density (b) Power conversion efficiency for both the QDSC.

### 4.3 Optimization of HTL

Appropriate concentration of acceptor dopants and thickness of HTL are crucial factors governing the performance of solar cell [31]. The best HTL is one that has the ability to reduce unwanted electrical resistances and combat recombination setbacks, thereby playing a key role in enhancing the device's functionality [31]. For this optimization, we have considered two aspects- optimizing the HTL thickness and HTL acceptor doping concentration.

#### 4.3.1 Comparative study of effect of HTL thickness in both the QDSC

To optimize the thickness of HTL, the HTL thickness is varied in increments of 10nm from 100-150 nm and the alteration in short circuit current density corresponding to the same is recorded. Initially the HTL thickness was taken to be 150 nm. The presumed thickness and doping concentration used is based on previously conducted experimental as well as simulating works using such materials. The impact of HTL thickness on short circuit current density of both QDSCs is illustrated in figure-7.

In case of both the quantum dot solar cells it has been observed that with a rise in HTL thickness, the photovoltaic current density increases linearly. As the thickness is raised from 100 -150 nm, for 50nm increase in thickness- the $J_{sc}$ increased from $20.625$-$20.626\ mA/cm^2$ for the $Sb_2S_3$ QD based solar cell and from $26.1906 - 26.1907\ mA/cm^2$ for the GaAs QD based solar cell. The change in efficiency recorded is of the order of four decimal places which shows that the impact of HTL thickness is not that significant enough to enhance the solar cell performance. Thereby we fixed the CuI-HTL initial thickness of 150 nm in both the QDSC.

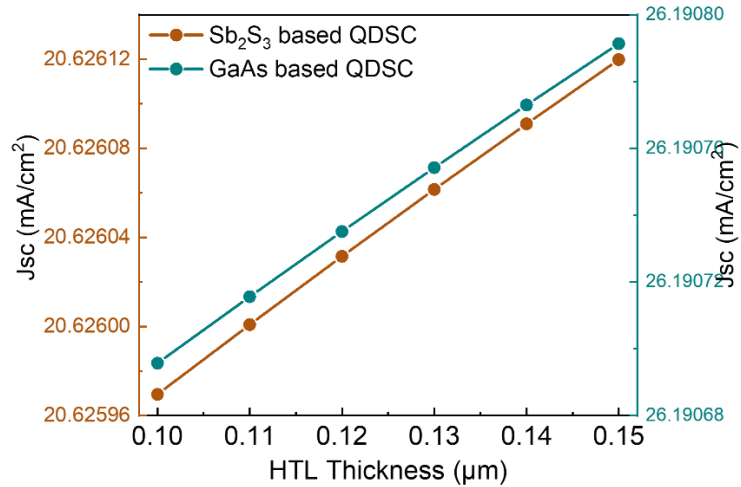

**Figure 7** Comparative study of effect of HTL thickness of the two QDSC on short circuit current density

### 4.3.2 Comparative study of the impact of concentration of acceptor doping in the HTL in both QDSCs

A solar cell device performance depends largely on HTL's acceptor concentration by varying the intensity of electric field at the HTL-absorber layer interface. A rise in electric field results in a larger separation of electrons and holes, thereby improving the solar cell efficiency [31]. The analysis examined the impact of concentration of acceptor doping in HTL ranging from $10^{15} - 10^{19}$ $cm^{-3}$ in five equal intervals on a logarithmic scale. In the SCAPS-1D simulation device, this kind of analysis is done using the batch setup function [8]. This function has the advantage that the user need not change the parameter to be studied manually, rather the user can input the range to be taken under consideration and the number of observations to be made within that range. The batch setup runs that many number of observations and generate the information, which is then used for the analysis.

The effect of concentration of acceptor dopants on (a) photovoltaic current density and (b) power conversion efficiency of the two QDSCs is depicted by figure 8. It has been observed that being the same material in the HTL, the trend followed is nearly same. With increase in HTL acceptor concentration, the current density decreased in both the solar cell. The fall in case of $Sb_2S_3$ QDSC is faster as compared to the GaAs QDSC. However, when analysed with respect to changes in magnitude of current density, it is found that the decrement is very low of the order $10^{-4}$ mA/cm². However, when the PCE is studied, it is found that the efficiency rises with rise in HTL acceptor doping concentration. However, the increase is about 0.1% for the $Sb_2S_3$ based solar cell whereas about 0.35% for the GaAs based solar cell. The dominating nature of GaAs is again reflected here. Although the HTL is same in both the QDSC, the different QD used makes the solar cell performance different from one another. Hence, we fix $10^{19}$ $cm^{-3}$ as the optimised acceptor concentration of the HTL.

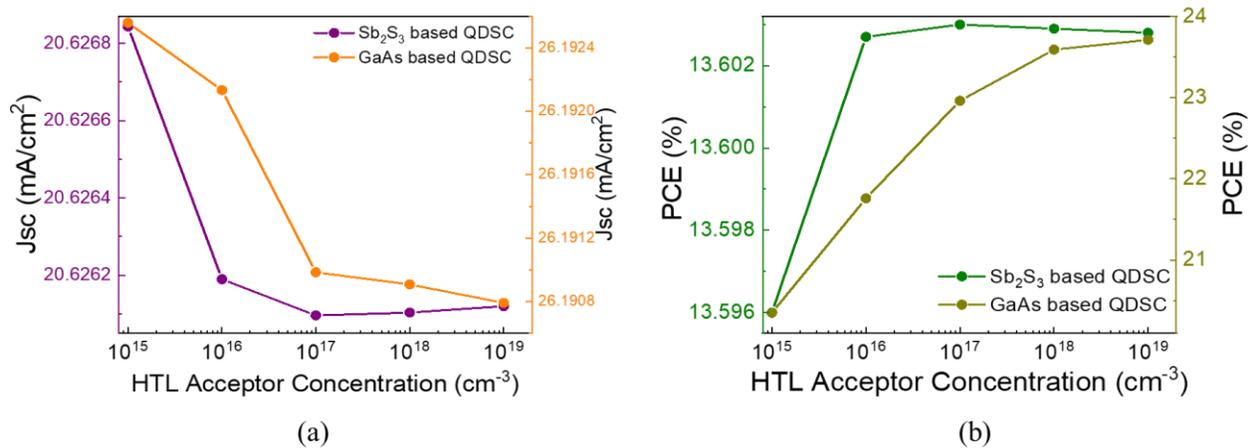

**Figure 8** Comparative study of impact of concentration of acceptor dopants in HTL of the two QDSC on (a) Current Density (b) Power conversion efficiency.

### 4.4 Impact of HTL –absorber layer interface defect density in the two QDSC

A solar cell efficacy is also influenced due because of the presence of imperfections at the interface of two adjacent layers of a solar cell [32]. At the junction of two adjacent layers, the probability of finding electron is the highest. Therefore, at these junctions, it is very likely that electrons may get trapped which results in low current generation and considerably lower efficiency [33].

Manipulation of the density of defects at the HTL- absorber layer interface in both the QDSC along with it a comparative analysis of the same has been conducted. The interface defect is made to vary from $10^{14} - 10^{18}\ cm^{-3}$ at the CuI-Sb$_2$S$_3$ interface and at the CuI-GaAs interface.

Figure-9 illustrates the relationship of density of interface defects with (a) PV-current density (b) open circuit voltage (c) fill factor and (d) power conversion efficiency.

It has been observed in both the QDSC that with the rise in density of interface defects, solar cell performance is negatively impacted. Figure 9 (a-d) is the clear representation of the above fact. As far as the efficiency is concerned, the decrement of the GaAs QD based solar cell shows sharp decrement with increase in HTL-absorber layer interface defect density as compared to the Sb$_2$S$_3$ QD based solar cell. The observed behaviour can be ascribed to the decrease in band gap of the absorber layer material as defect states are introduced due to their occupancy in the forbidden energy region, lowering the band gap [34]. Therefore, seeing the trend of decreasing interface defect density, we prefer to set $10^{14}\ cm^{-3}$ as the optimal density of interface defects at the HTL-absorber layer interface.

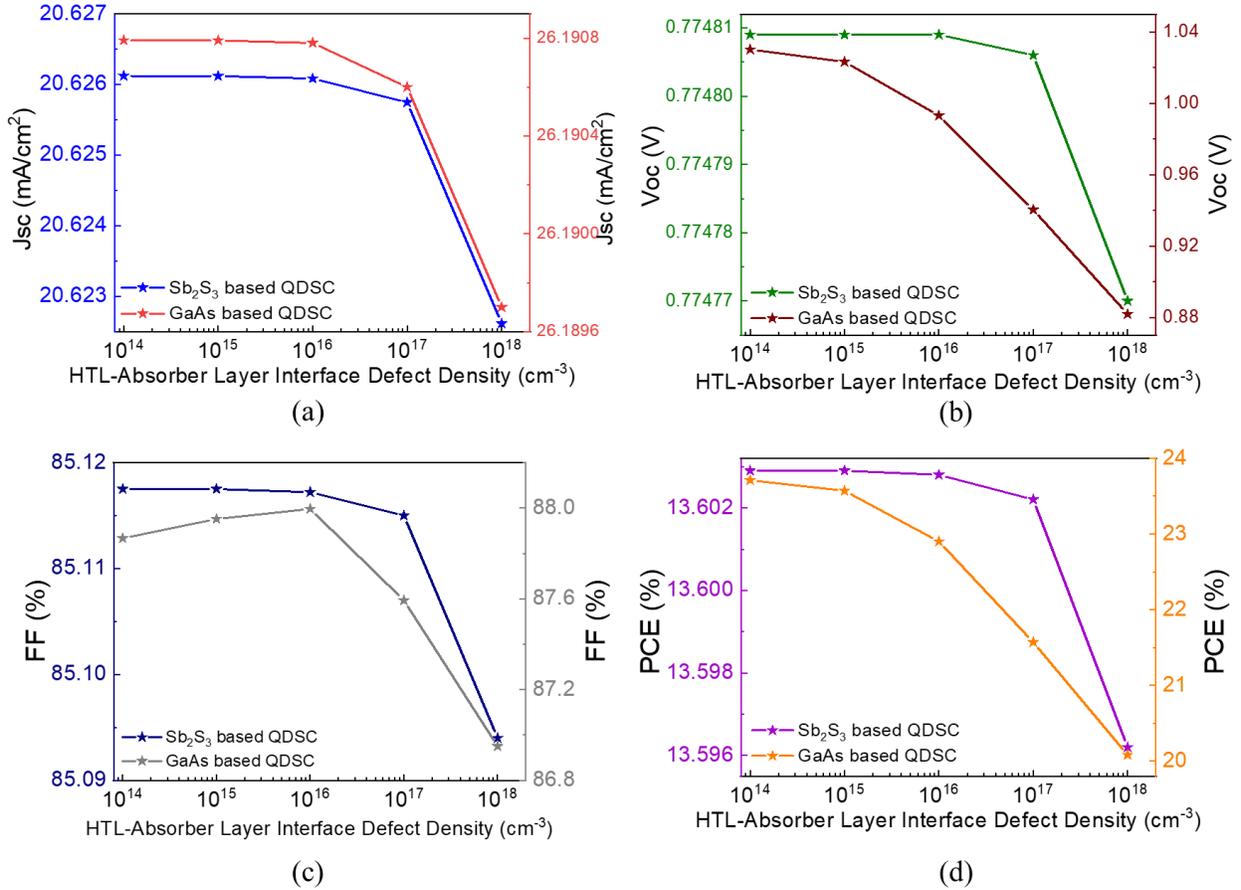

**Figure 9** Comparative study of the effect of HTL-absorber layer interface defect density of the two QDSC on (a) PV current density (b) Open circuit voltage (c) Fill factor (d) Power conversion efficiency.

## 4.5 Comparative analysis of the two optimized QDSC

Based on the studies carried out so far, we have optimized both the QDSC. The parameters that were utilized so to optimize the QDSCs are listed in table-2.

**Table-2:** Parameters that are used to optimize both the configurations of QDSC:

| Parameters | FTO | TiO$_2$ | CdS | Sb$_2$S$_3$ | GaAs | CuI |
|---|---|---|---|---|---|---|
| **Thickness (t) (µm)** | 0.350 | 0.050 | 0.020 | 0.550 | 0.550 | 0.150 |
| **Donor Concentration ($N_D$) (cm$^{-3}$)** | $2 \times 10^{19}$ | $1 \times 10^{19}$ | $1 \times 10^{17}$ | 0 | 0 | 0 |
| **Acceptor Concentration ($N_A$) (cm$^{-3}$)** | 0 | 0 | 0 | $1 \times 10^{19}$ | $1 \times 10^{19}$ | $1 \times 10^{19}$ |
| **Defect Density (Nt) (cm$^{-3}$)** | $1 \times 10^{14}$ | $1 \times 10^{14}$ | $1 \times 10^{14}$ | $1 \times 10^{14}$ | $1 \times 10^{14}$ | $1 \times 10^{14}$ |

The SCAPS-1D software provided us with the following information regarding the performance of the two optimized QDSC mentioned in Table-3.

**Table-3:** Result of simulation using the parameters mentioned in table-2 replacing the initial values mentioned in table-1:

| | **Sb$_2$S$_3$ QDSC** | | **GaAs QDSC** |
|---|---|---|---|
| **V$_{oc}$** | 0.8664 V | **V$_{oc}$** | 1.1241 V |
| **J$_{sc}$** | 22.15 mA/cm$^2$ | **J$_{sc}$** | 27.94 mA/cm$^2$ |
| **FF** | 85.89 % | **FF** | 88.69 % |
| **PCE** | 16.48 % | **PCE** | 27.86 % |

Table-3 shows that the GaAs QDSC outperforms the Sb$_2$S$_3$ QDSC in terms of effectiveness in solar energy conversion into usable electrical energy.

Figure-10 illustrates the comparative study of (a) J-V characteristics (b) Quantum Efficiency spectra of the two optimized QDSC.

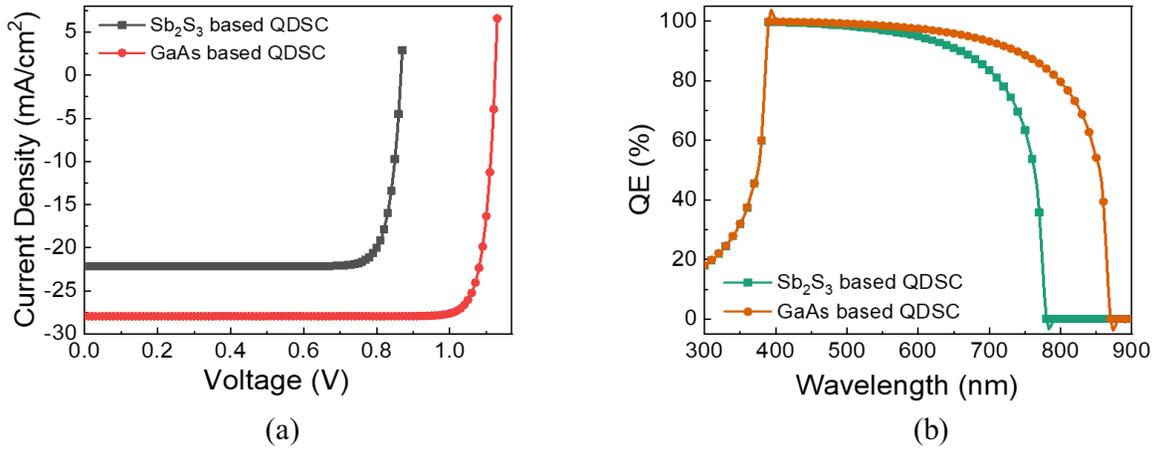

**Figure10** (a) J-V characteristics (b) Quantum Efficiency spectra of the two optimized QDSCs

## 4.6 Effect of solar cell operating temperature in two different configurations of QDSC

Temperature in the surroundings is not constant, it keeps changing depending whether it is day or night, due to weather conditions in an area or due to climate of a particular country [8], [35]. The working principle of a solar cell would remain the same whichever country maybe it is where it is working. Hence it is important to know, how the performance of the QDSC is affected due to change in temperature- which help us decide if there is any constraint in its application to any working environment. To understand this, we would vary the working temperature of the QDSC from 270 K (-3°$C$) to 320 K (47°$C$). It has been noticed that the solar conversion efficiency linearly diminishes with the rise in temperature. This is exclusively attributable to the fact that thermal expansion influences the size and structure of QDs. At the same time at low temperature, electrons being at rest possesses minimum energy. However, with the gradual rise in temperature, electrons are no longer at rest and they show a tendency to recombine with holes that is responsible for the reduction in solar cell efficiency with the gradual rise in temperature of the ambient [36].

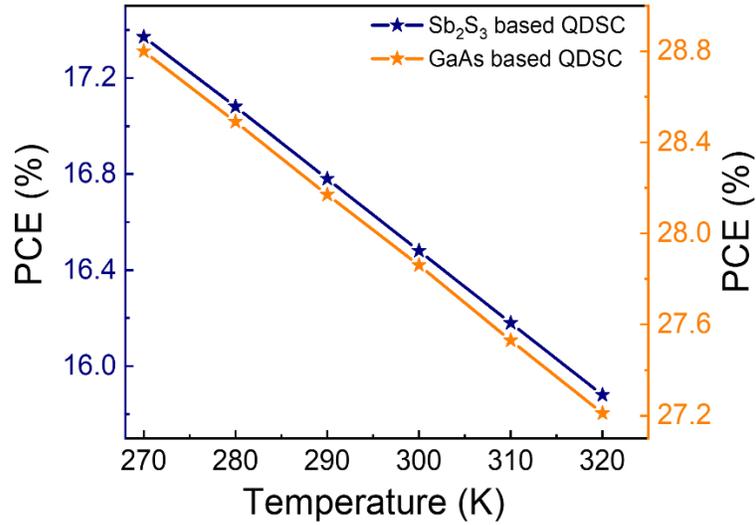

**Figure 11** Operating temperature effect on both the QDSC

## 4.7  Choice of back contact

Various back contacts can be used. Those most commonly used are Carbon with work function 5.02 eV, Nickel 5.22 eV, Gold 5.30 eV and Platinum 5.63 eV [12], [13], [14]. To check their effect the solar cell performance is checked with these back contacts and their performances are noted. It has been seen that not much efficiency is increased when we have used Ni, Au or Pt back contact instead of Carbon. Negligible increment in power conversion efficiency is noticed. By changing the back contact from carbon (5.02 eV) to Pt (5.63 eV), efficiency increased by 0.0001% in case of both the solar cells. But the cost incurred to increase the efficiency by 0.0001% is too high. Out of all the back contacts mentioned, carbon is the one found at a very low cost, therefore, thinking from the economy point of view, we have rejected the other back contacts and Carbon remained our choice of back contact for both the QDSC.

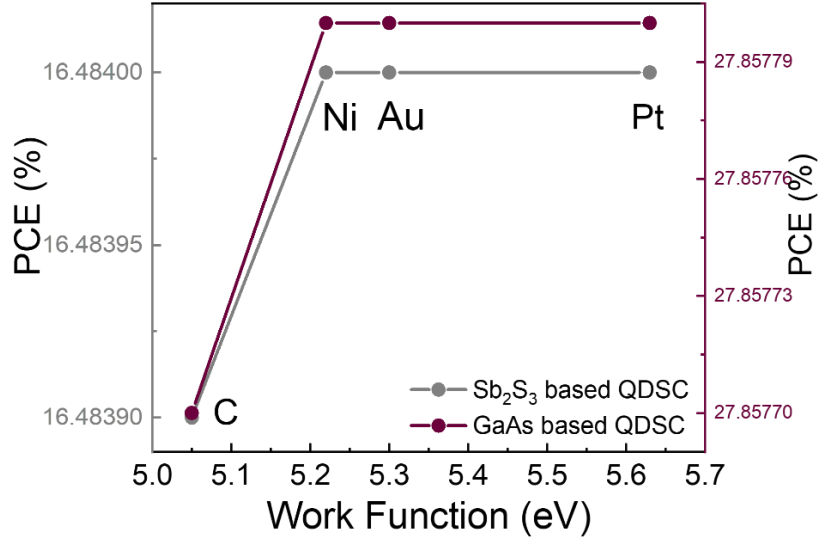

**Figure 12** Choice of back contact

## 4.8  Effect of parasitic resistances

While the solar cell is in a mode of operation, power dissipation occurs within the internal resistances, thereby lowering the efficiency of solar cells [37]. An ideal solar cell is characterized by zero and infinity series and shunt resistances respectively [38]. However, practically there do exist a finite non zero series and shunt resistances which tends to affect the solar cell performance. The shunt resistance is likely to arise due to leakage in p-n junction caused due to crystal imperfections [39]. So, here we have tried to deal with the way of optimizing the series and shunt resistances, so to extract the maximum possible power from the QDSC. On successful completion of studying this factor, we have also understood what is the maximum power conversion efficiency of the two QDSC. So, to investigate the influence of parasitic series and shunt resistance, we have systematically adjusted the series resistance within the range of 1- 6 $\Omega$ cm². When this was done, it was observed that the efficiency decreased in a linear fashion with the increase in the series resistance. Thereby, we have set 1$\Omega$ as the optimized series resistance. Now keeping this parameter fixed, we have adjusted the shunt resistance within the range of 1000-6000 $\Omega$ cm² in order to check the combined effect of both the resistances. It has been observed that the PCE increased and then saturated with the continuous increase in resistances. The PCE decreased from 16.48% to 15.94% in case of the $Sb_2S_3$ QDSC and from 27.86% to 26.95% in case of the GaAs QDSC when both the effects were taken into consideration. Thus, there is a net 0.54% and 0.9% loss in power conversion efficiency in $Sb_2S_3$-QDSC and GaAs QDSC respectively due to the combined effect of these resistances.

The impact that the parasitic resistances have on the efficiency of the solar cell is illustrated in figure-13 (a) and (b) and their corresponding contour plot in figure-13 (c) and (d).

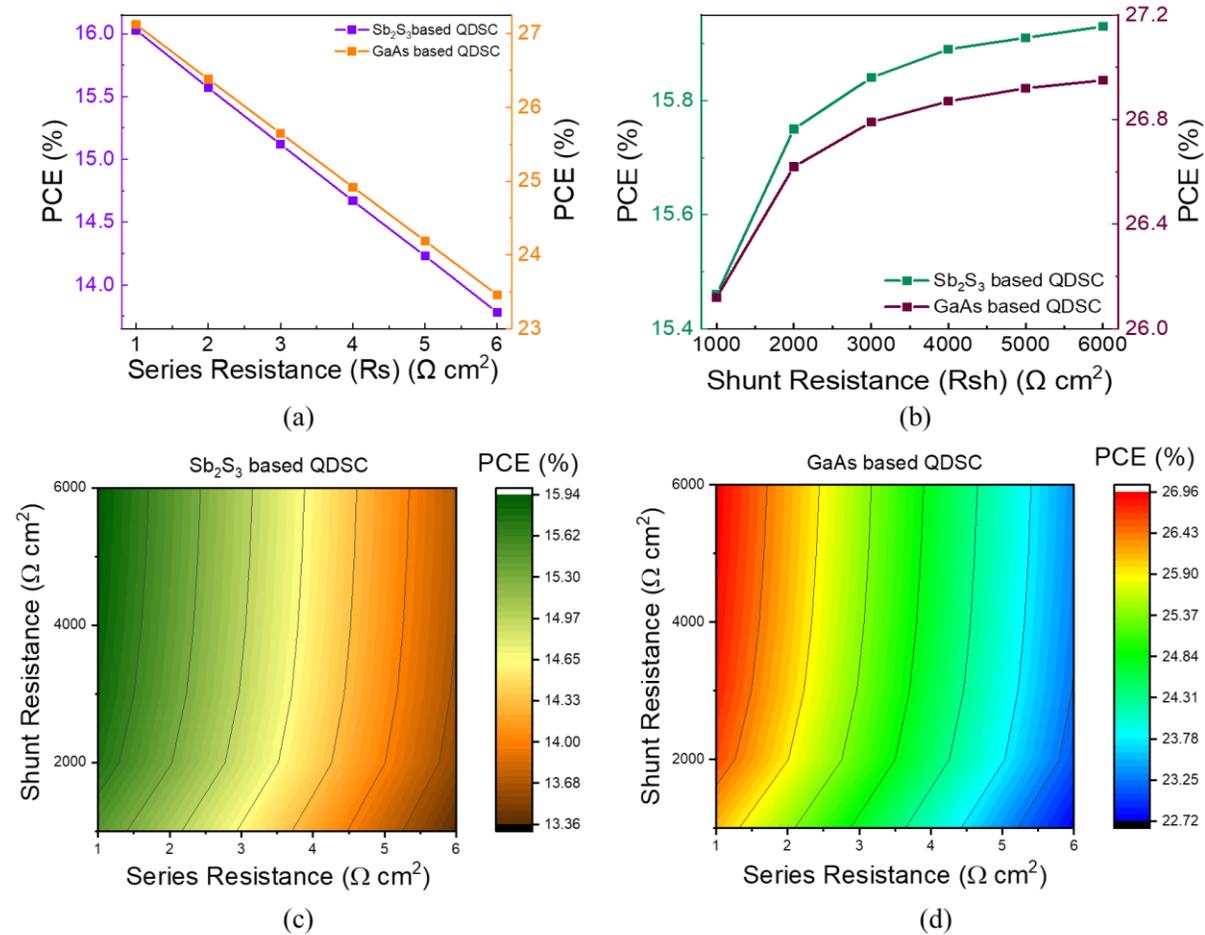

**Figure 13** Impact of parasitic (a) Series (b) Shunt resistances on PCE of both QDSCs; Contour plot representing impact of series and shunt resistance on PCE of (c) $Sb_2S_3$-QD (d) GaAs-QD based solar cell

## 4.9 Performance analysis

From the above performed studies, it has been noticed that the solar cell with GaAs-QD as the absorber layer has higher efficiency than the $Sb_2S_3$-QD based solar cell. The reason being - the material used in the absorber layer; GaAs is a direct band gap semiconductor whereas $Sb_2S_3$ possesses an indirect band gap. It has been found that in a direct band gap semiconductor, the recombination rate of its carriers is lower as compared to the indirect bandgap semiconductors [40], thereby giving better efficiency. The np heterojunction formed between the buffer layer and the absorber layer also affects the device performance. If the band energy alignment is not proper then it tends to increase the chances of recombination loses and the device performance degrades [41]. In case of $CdS/Sb_2S_3$ heterojunction, there do exist an imperfect energy level alignment, which is further responsible for the low efficiency of $Sb_2S_3$ QDSC [41]. Figure 1 (c-d) and figure 14 (c-d) confirms the above stated fact. As seen in figure 14 (c), there is a steep difference between the band energy alignment of CdS and $Sb_2S_3$, as compared to the CdS-GaAs heterojunction.

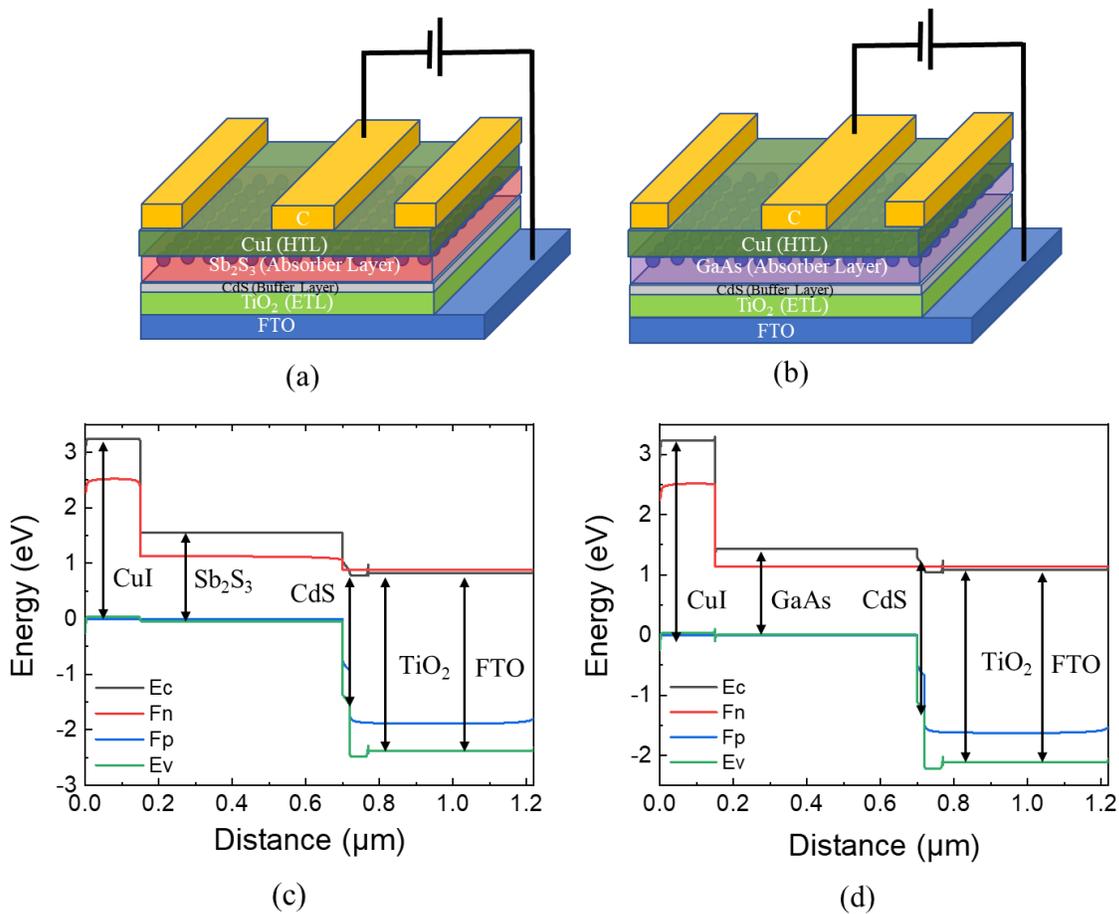

**Figure 14** Schematic diagram representing the solar cell device structure of the QDSCs (a-b) and their corresponding energy band diagram (c-d) defending the better performance of GaAs QD solar cell

## 5. Conclusion

In the present era for the designing of solar cells, numerical modelling has become an indispensable tool [26], [42]. In this work our aim is to identify a better absorber layer material and in turn identify the more efficient solar cell. The detailed study gave us the knowledge that the GaAs QD based solar cell is far ahead of the $Sb_2S_3$ QD based solar cell. The former achieves an optimized efficiency of 26.95% in presence of finite series and shunt resistances, whereas the latter could achieve an efficiency of only 15.94%. Since the device structure remains unchanged with the exception of the absorber layer, the characteristic trend one showed is not very different to that of the other. The band energy alignment contributes to the exceptional performance of the GaAs QDSCs. It has been established that efficiency shares inverse relationship with increase in temperature. This is a serious issue which must not be overlooked. In the tropical region countries where temperature goes quiet high would then not be able to produce better results. Since resistances significantly affect the solar cell's performance, attention must be focussed on reducing them. It has been reported that using composite phase change materials such as paraffin jelly in combination with perlite helps to reduce solar panel temperature and thereby contribute to maintain a higher efficiency [43]. The GaAs solar cell's efficiency dropped from 27.86% to 26.95% when it was subjected to the combined effects of series & shunt resistances.

## Declaration of Competing Interest:

The authors declare that they have no known competing financial interests or personal relationships that could have appeared to influence the work reported in this paper.

## Data availability:

Data will be made available on request.

## Acknowledgements:

The authors are thankful to the Indian Institute of Technology Jodhpur (IITJ) for the infrastructure facility.